# Robust and fast heart rate variability analysis of long and noisy electrocardiograms using neural networks and images

Sean P Parsons & Jan D Huizinga

*Farncombe Research Institute, McMaster University, 1280 Main Street West, Hamilton, Ontario L8S 4K1, Canada*

Heart rate variability studies depend on the robust calculation of the tachogram, the heart rate times series, usually by the detection of R peaks in the electrocardiogram (ECG). ECGs however are subject to a number of sources of noise which are difficult to filter and therefore reduce the tachogram accuracy. We describe a pipeline for fast calculation of tachograms from noisy ECGs of several hours' length. The pipeline consists of three stages. A neural network (NN) trained to detect R peaks and distinguish these from noise; a measure to robustly detect false positives (FPs) and negatives (FNs) produced by the NN; a simple "alarm" algorithm for automatically removing FPs and interpolating FNs. In addition, we introduce the approach of encoding ECGs, tachograms and other cardiac time series in the form of raster images, which greatly speeds and eases their visual inspection and analysis.

INTRODUCTION

A person's heart rate varies constantly, even when they are at rest. The reason for this is that the autonomic nervous system (ANS), which acts as both the break and accelerator on the heart, is also the central hub in a web of interactions between organs that define the "systems" physiology of the whole body [1, 2]. The feedback and cross-talk across the ANS make the heart rate an indicator dial of not just the heart's physiological state but also that of the whole body. It is perhaps the ultimate "biomarker", recognised by the fact that "heart rate variability" (HRV) is a large and flourishing field of inquiry [1]. But HRV not an easy biomarker to read.

At first sight much of heart rate variability appears chaotic and perhaps it truly is, in the mathematical sense. This is not surprising as the heart is picking up so many signals from all over the body. Sometimes one signal does become dominant, resulting in a more regular and identifiable pattern of variation. The cardiovascular system is a highly integrated part of the ANS and the kinetics of the interactions between its elements - blood pressure, heart rate and breathing - results in two resonant frequencies where those elements oscillate together [2, 3]. These frequencies are at ~ 0.1 and 0.25 Hz (6 and 15 cpm) and called Mayer waves and the respiratory sinus arrhythmia (RSA), respectively. By resonant it is meant, that driving the system at those frequencies, most easily by breathing at them, will produce the largest oscillations.

The fundamental datum of HRV is the tachogram - the time series of the heart rate or its inverse, heart beat (cycle) length. Each cycle consists of a regular series of peaks and troughs - the PQRST complex as first defined by Einthoven over a century ago. In principle any one of those peaks or troughs can act as the reference or "fiducial" point from which to calculate cycle length, but in practice it is almost always the R peak. The R peak is the largest and sharpest peak and so is the most localised in time and the easiest to detect by automated algorithm. The literature on R detection is vast. Often the algorithms are called "QRS" detectors as they rely on the signal characteristics about the R peak. Possibly hundreds of QRS detectors have been described that have used varying combinations of standard signal processing techniques - filters (linear and nonlinear), transforms (derivatives, wavelets) and thresholds and decision algorithms [4, 5]. Yet for all these algorithms, one of the very first, the Pan-Tompkins (PT), is still the gold (and commercial) standard thirty years after its invention [6]. What then has been the driving force of the multitudinous QRS detectors?

ECG recordings can suffer from a number of noise sources. The time scales (frequency components and sharpness) of most noises are distinct enough from the QRS that they present no hindrance to peak detection. If the noise has a lower frequency, for instance breathing oscillations and baseline drift (< 0.2 Hz), the QRS derivative will much higher and so can be thresholded for detection. If the noise frequency is much higher, for instance AC power interference (50 or 60 Hz), it can be filtered out (FFT, wavelet or empirical mode) without significant distortion of the QRS. However if the time scale of the noise is similar to the QRS, then their derivatives will likely be similar and designing a filter to separate them will be difficult. Electromyographic noise, resulting from electrical events in skeletal muscle, is the main contender in this category. A systematic performance comparison of several QRS detectors [7] showed that most of them detected close to 100 % of QRS embedded in low and high frequency noise, artificially generated to simulate drift, breathing and AC. But in the presence of noise with a wide band of frequency components, simulating electromyographic noise, their detection rates fell below 90 %. Noise then might be the driver behind the multiplication of QRS detectors. If there had been one outstanding algorithm in this regard, they would have stopped inventing them and the PT would be an historical footnote rather than the standard. Such an algorithm may however be on the horizon. It won't be a traditional signal processing algorithm but a neural network.

Neural networks (NNs) are a network of simple computing elements, modelled somewhat on biological neurons [8]. They have been around for decades, but is only in the last few years that a combination of computing power, large data sets and some theoretical breakthroughs have resulted in an ongoing revolution, with NNs applied both academically and commercially to just about every problem (and marketing opportunity) imagined by humankind. This revolution has variously come under the heading of "machine learning", "artificial intelligence" and "deep learning". The power of this revolution comes from the fact that NNs do in a sense learn. Rather than being designed with a very specific data set in mind, a single NN can complete a task (*optimisation*) with *any* data set it is first given examples for, in a process called *training* (i.e. learning). That is, it can generalize. There are general classes of NN architecture, suited for particular classes of task, but a single NN can carry out that task on any input data. For instance, a single convolutional NN can be trained to recognise in an image, dogs as well as cats or anything else you can imagine or photograph. A single reinforcement NN can learn to play video pinball as well as space invaders [9]. All that one needs to do is provide more training data for the extra thing you want the NN to do, not change the NN's actual architecture.

In contrast to NNs, traditional signal processing algorithms have poor generalization. No one has ever designed a combination of filters and transforms that can differentiate between cats, dogs, tables and chairs, or pick out QRS embedded in a wide variety of noises. Signal processing algorithms are typically designed with one dataset in mind and the addition of more dataset requires the addition of more steps - a change in architecture. The addition of each step requires a lot of empirical effort for integration, to keep the overall algorithm sensitive and stable. It is this fact that must explain the nonarrival of that outstanding algorithm that will surpass the PT.

What then of NNs? The literature on NNs in ECG analysis goes back at least to the early 1990s but most of this is not related to actual QRS detection. NNs have been employed in three roles:

1) *Convolutional Classification*. CNNs were invented for image recognition, but can just as easily be applied to one dimensional signals such as ECGs. CNNs do not need to be told first where in an image the object to be recognised is. They do this implicitly - a property called translational invariance [8]. Similarly CNNs can be trained to recognise waveforms in an ECG, such as might be correlated with a pathology, without being giving the location of particular cycle markers - i.e. without QRS detection. The CNN is trained with segments of a few seconds to minutes, labeled according to the arrythmia or other pathology recognised in the patient [10, 11]. The CNN is only as accurate as the trained cardiologists used to label the data (i.e. pretty accurate), but has the advantage that it can represent the learned opinion of a cohort of cardiologists rather than just one and works much faster and for less pay. CNNs are now the backbone of a number of commercial products, for both the clinical and recreational market (e.g. Alivecor, Cardiologs, Mawi, PocketECG, Cardiogram).

2) *QRS detection*. A CNN architecture called an encoder-decoder can be used to return the position of an object in an image [8]. We are not aware that such an architecture has been used for QRS detection. The alternative to an encoder-decoder for locating the position of an object, is the sliding window approach (sometimes called pixel-based classification). A segment of the dataset (small patch of an image or short segment of a times series) is passed to the NN and the NN's job is to say whether the object is present at some fixed point in that window. By sliding the window over the dataset (image or ECG) one can find the position of all the objects. A few authors have used this approach for QRS detection [12-16].

3) *QRS classification*. A traditional signal processing algorithm (filter or transform) is used to detect candidate peaks and then an NN is used to verify them as P/R/T/etc or to classify their morphology in relation to pathology. This approach has been used by a number of authors [17-26]. Given that any NN that can classify a peak can also locate it, by using a sliding window, this two-part process is rather surprising. It is more computationally efficient (fast) than the sliding window, but this is less a concern now than it was twenty years ago.

4) *Noise Detection and Filtering*. NNs have been used as nonlinear filters for noise removal [27] and segmentation of noisy segments [28].

The QRS complex is so obvious in an ECG free of noise, that even a simple algorithm like the PT will give 100 % accuracy - 100 % sensitivity (true positive rate) and 100 % specificity (true negative rate). Therefore it is no surprise that the handful of studies that have used NNs for QRS detection (see above) all report upwards of 99.9 % (often 100 %) for these measures, because not one of them specifically tested noisy ECGs. They all used standard databases (such as the MIT/BIH) where noise-contaminates QRS make up a zero or nominal fraction of the total. None made any specific analyses of this fraction. With such data the performance of an NN is a forgone conclusion. In the present study our objective was rather less prejudiced. We wished to see how well an NN QRS detector would fair with noise and if there were limitations, what might be done about them. This objective was in the wider context of HRV analysis of ECGs of several hours length that were often far from clean.

METHODS

*Neural Network*

The neural network was built and trained with TensorFlow, Google's open source NN library. All coding was in a Jupyter notebook (Supplementary Data) using TensorFlow's Python API. Trained models were saved using the saved model builder, so that they could be accessed by CardioImage (see below) using TensorFlow's Java API.

The NN consisted of five layers of densely connected neurons (units) with ReLU activation functions. The first layer consisted of 80 units which took as input a 200 ms window of ECG (100 samples at a 500 Hz sampling frequency). In machine learning terms, each window is a 100 element feature vector. The layers then tapered down in size to a final "logit" layer of 2 units, which would activate in proportion to the probability that there was an R peak at the centre of the window (the 50[th] sample) or not, respectively. By sliding the window across the ECG, two corresponding activation signals were thus produced - "is an R peak" and "not an R peak" and R peaks were detected where the former was greater than the later (Fig. 2).

Table 1 | Structure and regularization of the neural network.

|  | layer | | | | |
| --- | --- | --- | --- | --- | --- |
|  | # 1 | # 2 | # 3 | # 4 | logit |
| no. neurons/units | 80 | 40 | 20 | 10 | 2 |
| drop out rate | 0.1 | 0.1 | 0.1 | 0.1 | - |
| batch normalisation momentum | 0.5 | 0.5 | 0.5 | 0.5 | - |

For regularization we used drop out and batch normalisation (Table 1). The loss function was softmax cross entropy. Training was with gradient descent at a learning rate of 0.01.

*Training Data and Protocol*

Training data was collated by a CardioImage plugin (Supplementary Data) from the ECGs of five subjects. It was encoded as $n_{feature}$ x $n_{window}$ images, where $n_{feature}$ was the number of samples (features) in a window and $n_{window}$ was the number of windows (training samples or instances). All windows were normalised as,

$$\mathbf{w} \rightarrow (\mathbf{w} - \mu)/\sigma \tag{1.1}$$

where w was the window (feature) vector and μ and σ were the mean and standard deviation, respectively, of a noiseless segment of its ECG.

Image banks were read in the Jupyter notebook for training the NN. The three banks were:

1) *R peaks*. 7651 windows with R peaks at their centre, from noiseless segments of ECG. The R peaks had been detected by PT.
2) *Background*. 10998 windows without R peaks at their centre, from noiseless segments of ECG. They were selected from centres randomly distributed between R peaks.
3) *Noise*. 7500 windows were selected at random from noisy segments of ECG where it was hard to see any R peaks.

The R peak and background banks are very homogeneous - one PQRST complex looks much the same as any other, both within a single ECG and between subjects. Therefore we augmented these banks by adding a sine wave to each window with a random frequency, amplitude and phase, plus white Gaussian noise with a random amplitude (Table 2).

Table 2 | Window augmentation.

|  | units | mean | S.D. |
|---:|---|---:|---:|
| offset | normalised amplitude | 0.0 | 0.2 |
| frequency | window$^{-1}$ | 0.5 | 0.5 |
| phase | window | 0.0 | 1.0 |
| amplitude | normalised amplitude | 0.1 | 2.0 |
| Gaussian white noise | normalised amplitude | 0.1 | 0.02 |

After an initial round of training we applied the model to full length ECGs with lots of noise which had already been passed through an R peak detection algorithm (PT or the NN) and then manually corrected to give a ground truth. The false positives detected by the NN were then used to create a *false positive* bank which could be used for further training, either from scratch or from the NN's previously trained state. In either case, the cycle of training and false-positive generation could be iterated, each time the number of false positives decreasing (see Results).

Training was performed on a fixed proportion of windows randomly sampled from each bank (Table 3).

Table 3 | Mix of training instances (windows) used.

| bank | no. windows |
|---|---|
| noisy background | 4500 |
| augmented background | 4399 |
| augmented R-peak | 4950 |
| false positive | 2961 |
| TOTAL | 16450 |

The complete set of windows (Table 3) was split into training and validation sets in the proportion of 9:1, respectively, by random sampling. For each epoch batches of 500 windows were sampled from the training set. To produce smooth training curves (false/true positive/negative rates), five models were trained from the same starting point and the training curves of these five models were averaged (Fig. 1). The last model was saved for testing and use in CardioImage (see below).

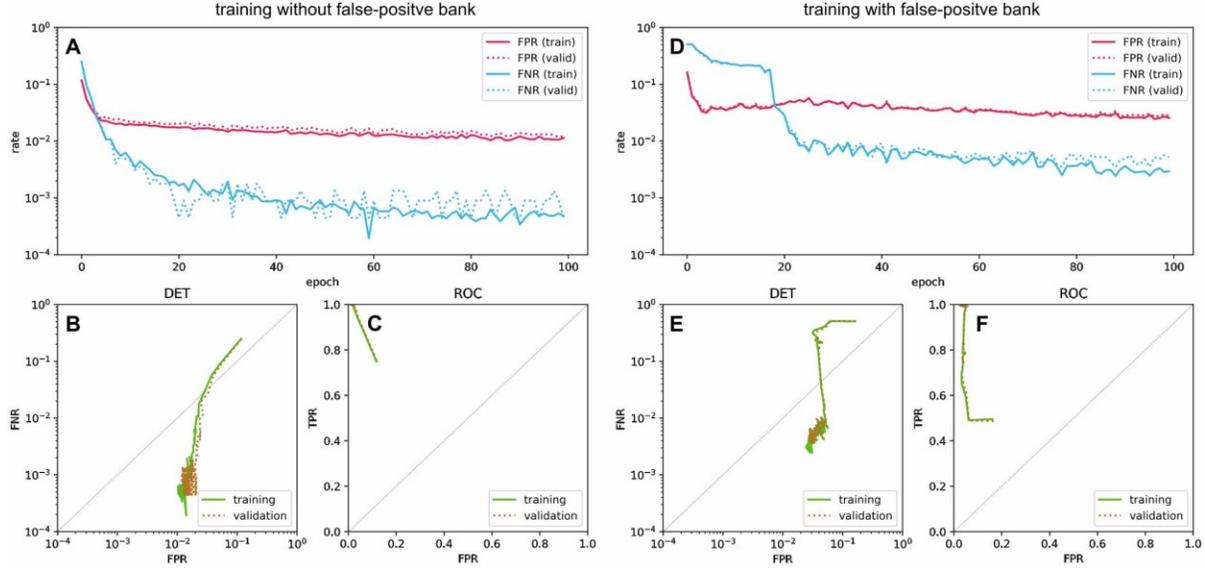

Figure 1 | Training curves of the NN.

A and D, The NN was shown the training data set (Table 3) multiple times and with each time or epoch, its false positive rate (*red*) and false negative rate (*blue*) decreased. FPR and FNR are the complements of selectivity and sensitivity, respectively - e.g. $10^{-2}$ FPR is equivalent to 99 % selectivity. In A, the NN was trained with the entire set minus the false-positive bank and in D it was trained with the entire set. Validation error rates (*dashed*) followed those of the training data (*solid*), showing that regularisation of the NN (drop-out and batch normalisation) were working. B and E, plots of FNR as a function of FPR (detection error tradeoff curves) over the same epoch range. C and F, true positive rate as a function of false positive rate (receiver operating curves) over the same epoch range.

*Diagnosis*

Given a sequence of R peak detections, each has a time ($\tau$), the RR interval preceding it ($\delta^-$), the RR interval following it ($\delta^+$) and the normalised absolute difference between those intervals ($\Delta$),

$$\begin{aligned}
\tau_k &= \text{time of k}^{th} \text{ detection} \\
\delta_k^- &= \tau_k - \tau_{k-1} \\
\delta_k^+ &= \tau_{k+1} - \tau_k \\
\Delta_k &= \left|\delta_k^+ - \delta_k^-\right| / \min\left\{\delta_k^+, \delta_k^-\right\}
\end{aligned} \qquad (1.2)$$

Commonly $\Delta$ is less than 0.2 for true consecutive RR intervals ($\delta$). At most, during a premature contraction $\Delta$ may rise to 0.5. Where there is a sequence of false negatives, a gap with undetected R peaks, then the $\Delta$ of the detections either side of this gap will be approximately equal to the the number of missed peaks. Conversely if there are one or more false detections between two true detections, the Δ of the latter will be near to one or higher. In this way Δ can be used to robustly detect false positives and false negatives as well as assess their likely number and sequence.

In CardioImage Δ can be plotted as a raster image time series, complementary to the ECG, or the Δ-δ co-distribution can be plotted as an image where each pixel is a Δ-δ bin (Fig. 6 E, H).

*Alarm Sequences and Interpolation*

The first task is to find detection sequences where Δ greater is than some threshold ($\alpha$). The first and last elements of a sequence are (usually) true detections, whilst in between there must be either missing (false negative) detections to interpolate or false positive detections to remove or both. Such a sequence we call an "alarm sequence".

To complicate matters there may be some false detections that by chance are spaced such that their Δ is less than $\alpha$. We may therefore wish to find sequences of $\Delta > \alpha$ that are not entirely contiguous, but may contain some gaps where $\Delta < \alpha$, each gap with a maximum size of $\beta_{max}$. To make this clearer, consider a detection sequence converted into binary according to whether $\Delta > \alpha$,

$$\Delta > \alpha = [00000011001111011100000] \qquad (1.3)$$

With $\beta_{max} = 0$, there are three alarm sequences [11], [1111] and [111]. With $\beta_{max} = 1$, there are two alarm sequences [11] and [11110111]. With $\beta_{max} = 2$, there is one alarm sequence, [110011110111].

Table 4 | Alarm sequence parameters used in this study.

| parameter | meaning | value |
| --- | --- | --- |
| $\alpha_0$ | initial Δ threshold | 0.6 |
| $\alpha_1$ | subsequent Δ threshold | 0.2 |
| $\beta_{max}$ | number of gaps to skip | 2 |

Furthermore we may want to use a higher threshold ($\alpha_0$) to find the first element in an alarm sequence, than for subsequent elements ($\alpha_1$), because of that chance of false positive detections just happening to be spaced so as to lower their Δ. The first ($u$) and last ($v$) elements of the alarm sequence are found as,

```
if Δu > α0:
    v = u + 1
    no. of gaps = 0
    while (v <= n) and (no. of gaps < βmax):
        if Δv < α1:
            no. of gaps += 1
        else:
            no. of gaps = 0
        v += 1
```

$$v \mathrel{-}= (\text{no. of gaps} + 1)$$

Once an alarm sequence is found, all detections between the first and last element are removed. The next task is then to interpolate a set of detections between the first and last element, such that the resulting intervals interpolate between $\delta_u^-$ and $\delta_v^+$. First we divide the period between the first and last elements ($\tau_v - \tau_u$) into the $m$ equal intervals nearest to the mean of $\delta_u^-$ and $\delta_v^+$. $m$ is found by,

$$\delta^{uv} = \tau_v - \tau_u$$
$$\delta^{int} = (\delta_u^- + \delta_v^+)/2$$
$$m = 1$$
while True:
  $a = |\delta^{uv}/m - \delta^{int}|$
  m += 1
  $b = |\delta^{uv}/m - \delta^{int}|$
  if a < b:
    break
m -= 1

If $m$ is greater than one, then detections need to be interpolated between $\tau_u$ and $\tau_v$. First we we initialise a vector ($\delta$) of $m$ intervals equal to $\delta^{uv}/m$. We then adjust the intervals iteratively so that they go from $\sim\delta_u^-$ to $\sim\delta_v^+$,

$$\delta = \delta^{uv}/m$$
$$a = \delta_u^- + \delta_v^+$$
$$j = 1$$
for $i$ = floor($m/2$) - 1 to 0, $i$--:
  $\delta_i \mathrel{-}= (j * a)$
  $\delta_{m-i-1} \mathrel{+}= (j * a)$
  $j \mathrel{+}= 1$

Detections are then interpolated according to $\delta$.

If $m$ is one and $v - u = 1$ (i.e. the alarm sequence is [11]), then either detection $u$ or $v$ must be false. The detection with the smaller interval is removed, so that the resulting intervals either side of the remaining detection are nearest to each other.

if $\delta_u^- < \delta_v^+$:
  $u$ removed
else
  $v$ removed

RESULTS

*A Neural Network for QRS Detection*

Our neural network was motivated by two observations:

1) ECG noise commonly presents in the form of sharp deflections that cannot be distinguished from R peaks by the PT or a similar signal-processing algorithm (Introduction). A neural net should easily be able to distinguish between the two by being trained with examples of both.

2) There are often long segments of intense noise when QRS cannot even be identified by eye (Fig. 2 A) and so are unlikely to be detected by an NN. In this case it would be better for the NN to detect nothing, generate false negatives (FNs), rather than be so sensitive as to give false positives (FPs). The gaps left by the NN could then be detected and interpolated by a second algorithm.

We constructed a NN of five densely connected layers that took as input a 200 ms window of ECG and output two values - the likelihood that there was or wasn't an R peak at the centre of the window (Methods). The window was slid across the ECG to give the corresponding two signals across time (Fig. 2 B) and where the "is-an-R-peak" signal was the greater, an R-peak was detected (Fig. 2 C).

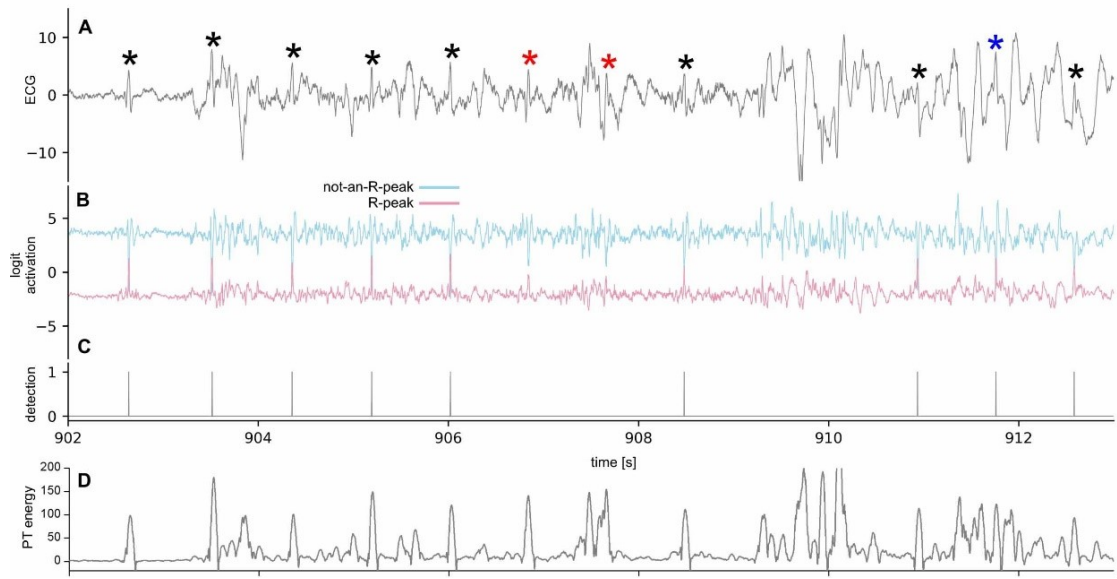

Figure 2 | R-peak detection during noise.

A, a segment of ECG contaminated by electromyographic noise. R peaks can be visually identified (*asterisks*) from their shape and timing, but there are presumable gaps where the R-peaks are lost in the noise. B, the neural network's final "logit" layer has two neurons - one which activates according to whether the NN thinks there is an R-peak (*pink*) and the other that activates according to whether the NN thinks there is *not* an R-peak (*blue*). When the former is greater than the latter, a detection is marked (C). Most of the R-peaks visible by eye are detected (*black asterisks*), but some are not (*red asterisks*) - the R-peak neuron activates, but not greater than the not-an-R-peak neuron. R-peaks cause peaks in the energy signal of the PT algorithm (D), but so does the noise.

Generally the NN was better than the PT algorithm at distinguishing QRS from electromyographic noise (Fig. 2 D). The NN could miss QRS that the PT spotted (Fig. 1 A, red asterisks) but this was the outcome of our objective to detect generate FNs, rather than be so sensitive as to give FPs (2, above). The NN was initially just trained with examples of: R-peaks (windows with R peaks at their centre); background (windows without an R peak at the centre); and noisy background (as before but with noise). When this model was used on a noisy ECG series it picked up R-peaks, but also a lot of FPs (Fig. 3 A - C). These FPs were automatically identified and collated from several noisy ECGs, by comparison to detection series (either by PT or NN), manually edited to remove them. The collated FPs, a "false-positive bank", were then added to the training set and the model trained again from scratch. Including the false-positive bank suppressed FPs, leaving gaps were there were segments of noise (Fig. 3 F and G).

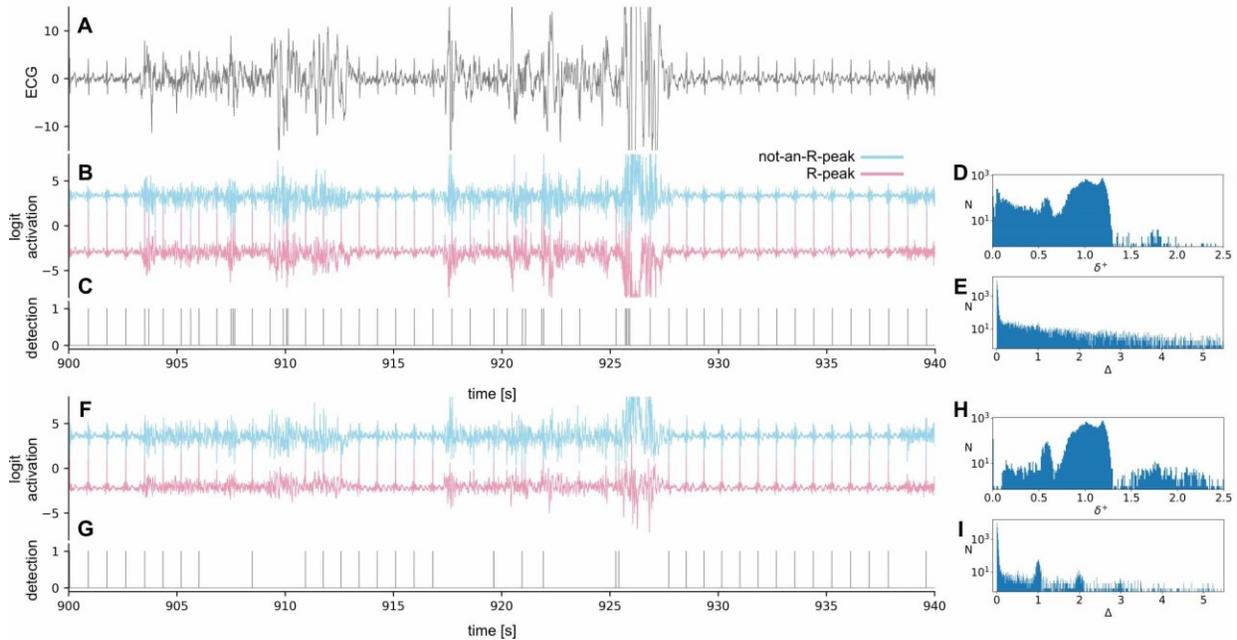

Figure 3 | NN detections without and with false-positive training.

A, a longer segment from the same ECG in Fig. 1 A, with two periods of electromyographic noise. B, the ECG was passed through the NN trained without false-positives. For each detection over the whole six hour ECG, the interval to the following detection ($\delta^+$) and the absolute difference between the preceding and following intervals, normalised to the lesser of the two ($\Delta$) were calculated. The distribution of $\delta+$ (D) has two humps at ~0.6 and 1.2 s, representing real RR intervals, on top of a broad sweep of d+ up to the second hump, representing false-positives. The latter gave a broad distribution of $\Delta$ (E). Training the same NN from scratch, but this time including a bank of false-positives, reduced the number of detections (F, G). This reduces the background across the $\delta^+$ distribution (H) and creates integer peaks in the $\Delta$ distribution, representing gaps with false-negatives (I).

To diagnose the performance of the NN without visually scanning through the whole trace, we measured two quantities for each detection: the interval following the detection ($\delta^+$) and the absolute difference of the intervals either side of the detection, normalised to the shorter of them ($\Delta$) (Methods). In a six-hour ECG run through the model trained without the false-positive bank, $\delta^+$ was broadly distributed between 0 and 1.3 s (Fig. 3 D). There were two humps at ~ 0.6 and 1 s reflecting true RR intervals (Fig. 3 D). $\Delta$ was broadly distributed between 0 and 5 (Fig. 3 E). These broad distributions can be understood from the point of view that FPs occurring at random will give random $\delta^+$ and $\Delta$. When the false-positive bank was included in training, the two humps in $\delta^+$ became more distinguished, with few detections in between (Fig. 3 H), and $\Delta$ was concentrated at peaks at 0, 1, 2 and 3 (Fig. 3 I). The integer peaks > 0 reflect gaps with that number of FNs. True RR intervals incremented or decremented by only small amounts, with $\Delta < 0.1$ (Fig. 3 I).

*Diagnosing and Interpolating*

$\Delta$ can be used for identifying sequences of one or more FPs and/or FNs - the $\Delta$ of the true detections (R peaks) at either side of the sequence will be large (Methods). Therefore identifying such "alarm sequences" is just a matter of thresholding $\Delta$. FPs can then be removed and the full period of the sequence divided into approximate "RR" intervals interpolated between the true intervals either side.

The alarm sequence detection and interpolation algorithm (Methods) was very robust (Fig. 4). Noisy segments of ECG were clearly demarcated by alarm sequences (Fig. 4 A and C). Interpolated detections were often close to NN detections that had been removed (Fig. 4 C), suggesting that the latter were true-positives. The alarm algorithm drastically changed the distribution of $\delta^+$ and $\Delta$. The peaks in $\delta^+$ were more distinct, with little background spread at the lower and higher, unphysiological ranges (Fig. 4 E).

Normal RR intervals gave $\Delta$ of less 0.3 (Fig. 4 G) which made selection of the alarm $\Delta$ thresholds (Table 4) straight forward and standard across ECGs - there was no need to tune. However premature atrial or ventricular contractions, gave $\Delta$ up to ~0.5 (Fig. 4 H-J). Whilst these physiologically normal events are fairly rare their frequency is not nominal in healthy subjects. By setting a relatively low threshold ($\alpha_0 = 0.6$), some would inevitably be removed by the alarm algorithm and have to be manually put back in (see below).

*CardioImage*

ImageJ is a popular open source image analysis software developed by the National Institutes of Health (USA). "Plugins", pieces of Java code, can be written by anyone to add functionality to the core software. We developed a suite of ImageJ plugins for calculation of tachograms and their analysis, collectively called CardioImage. The plugins implement our trained NN, a PT algorithm, alarm and interpolation algorithm, manual editing, and two common HRV analyses - FFT spectra and ensemble cycles.

ECGs, tachograms and other times series were encoded as row-major raster images - i.e. each time sample is represented as a pixel in a two-dimensional array of such pixels (an image), with time running from left to right across a row of pixels, then continuing along the next row and so on, to the bottom of the image. Multiple channels (for example impedance or NN output) were arranged as a stack of images. Viewing time series data as images has the advantage that the gross features of a long series can be visualised at a glance - for instance, the long-term variations in heart rate in a tachogram.

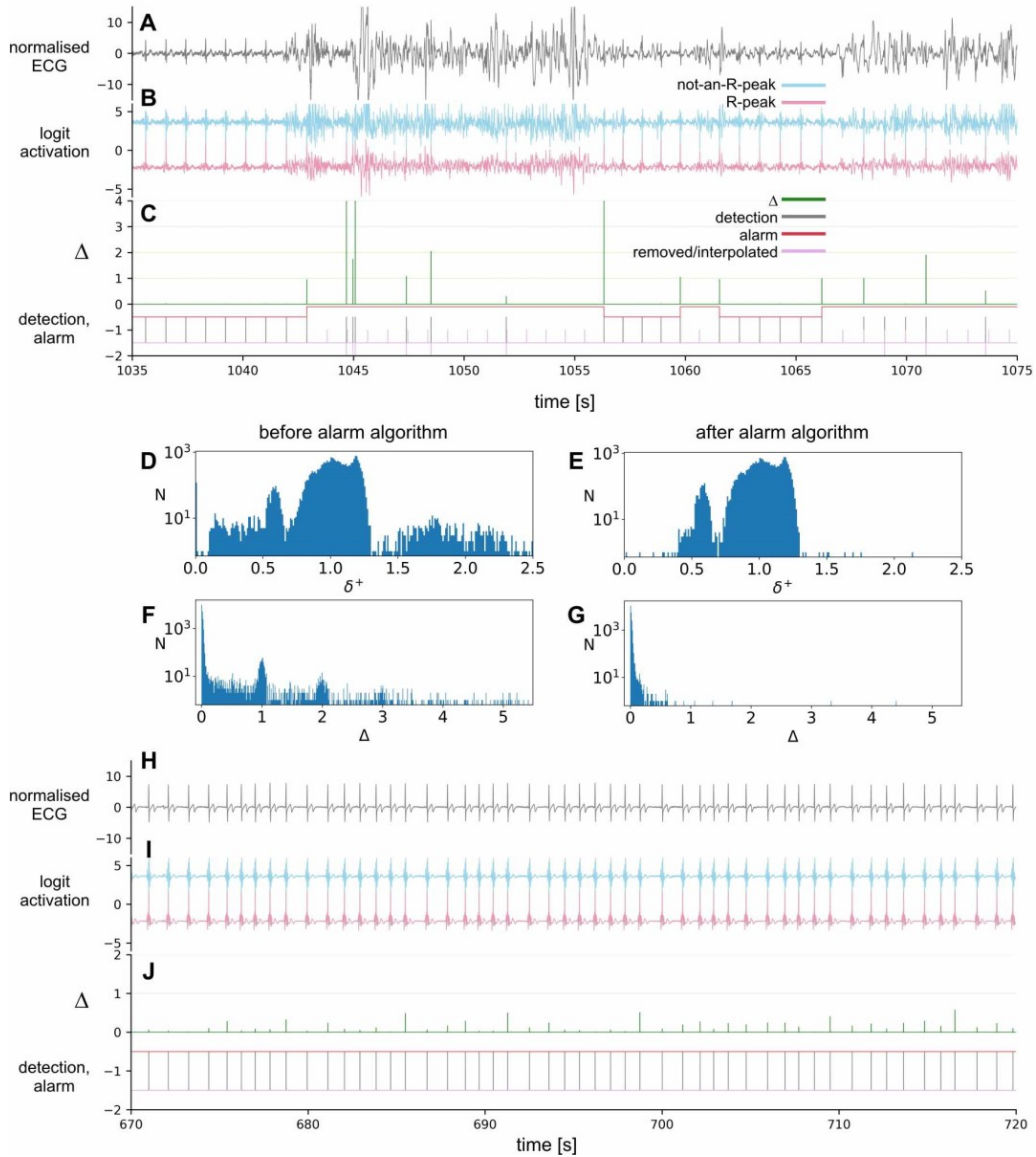

Figure 4 | Alarm algorithm

A, a segment from the same ECG as in Fig. 2 and 3. B, the ECG was passed through the NN trained with false-positives. C, each NN detection (*grey tick*) has a Δ value (*green tick*) which tend toward integer values (*horizontal green lines*) equal to the number of missed detections (false negatives) either preceding it or following it. Sequences of Δ greater than some threshold can be used to define an alarm sequence (*upward deflection in red line*; see Methods for details). Detections within this alarm sequence are removed (downward ticks in pink line) and new "detections" are interpolated (*upward ticks in pink line*) according to the intervals either side of the alarm sequence. This "alarm algorithm" removes spuriously short or long detection intervals ($\delta^+$; D, E) and Δ is restricted to values under ~0.3 (F, G). H-J, Δ can reach as high as 0.5 during premature atrial or ventricular contractions as shown in this example of the former (a different subject and ECG from A). The highest value of Δ is for the premature R-peak itself, with an abnormally short preceding interval ($\delta^-$) and an abnormally long following interval ($\delta^+$). The Δ of the preceding and following R peaks will also be abnormally large, but less so.

CardioImage is described in detail in the user guide (Supplementary Data). Here we outline it with a step-by-step pipeline example.

1) *Import.*

A Mindware binary file (.mw) is read into a raster image stack (Fig. 5). It has three channels (images) - ECG, impedance and impedance time derivative. We do not yet support other binary file formats, but any text file (one column per channel) can be imported.

2) *R peak detection.*

From the imported stack, the R peak plugin creates a second image stack ("detection stack") of four images/channels: the ECG; detection signal; detection markers; and $\delta^+/\Delta$ (Fig. 5). All but the ECG are initially blank. An NN is selected by referring to its build folder, created by the Python training code. In this way, any number of NN models, with differing architectures or training regimes, can be used. The only stricture is that the model must refer to variables for the window size, input tensor and logit layers, with a standard naming so that they can be identified by the plugin. The PT algorithm is also implemented.

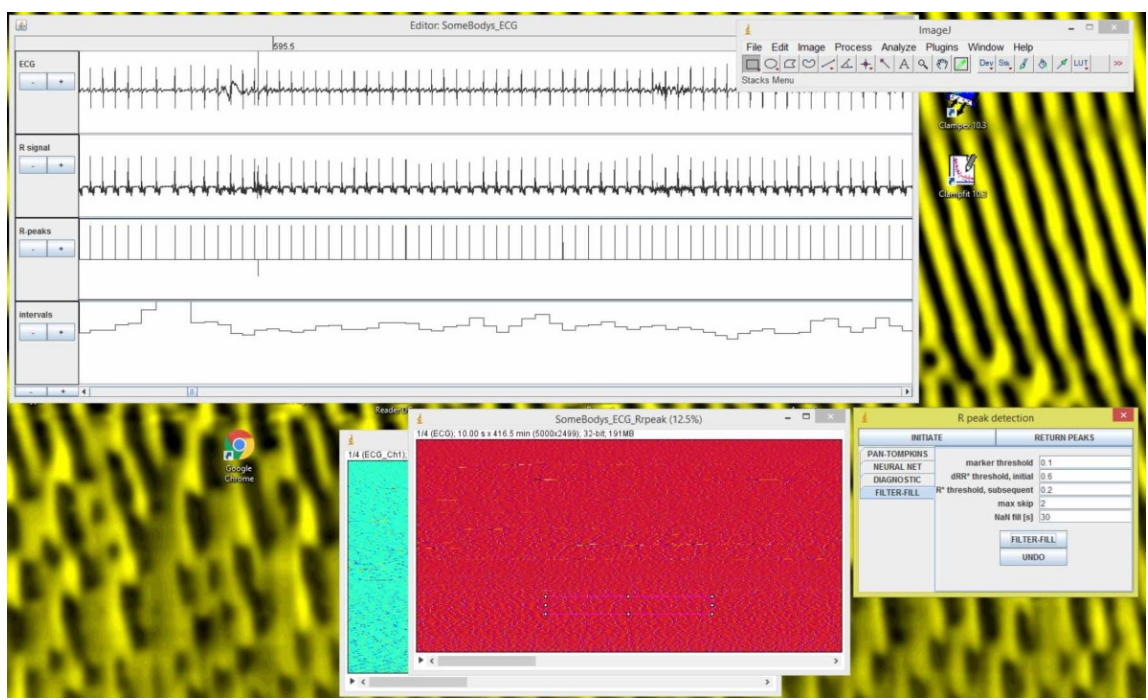

Figure 5 | R-peak detection plugin

The plugin takes an input stack of images containing the ECG data (lower middle window at back) and creates a second "detection stack" (lower middle window at front) containing four channels - the ECG, the detection signal (in this case the activation of the "is-an-R-peak" neuron), detection markers and RR intervals ($\delta^+$). A multichannel plot window (upper left) shows the same channels (*top* to *bottom*) and can be used for manually editing detections.

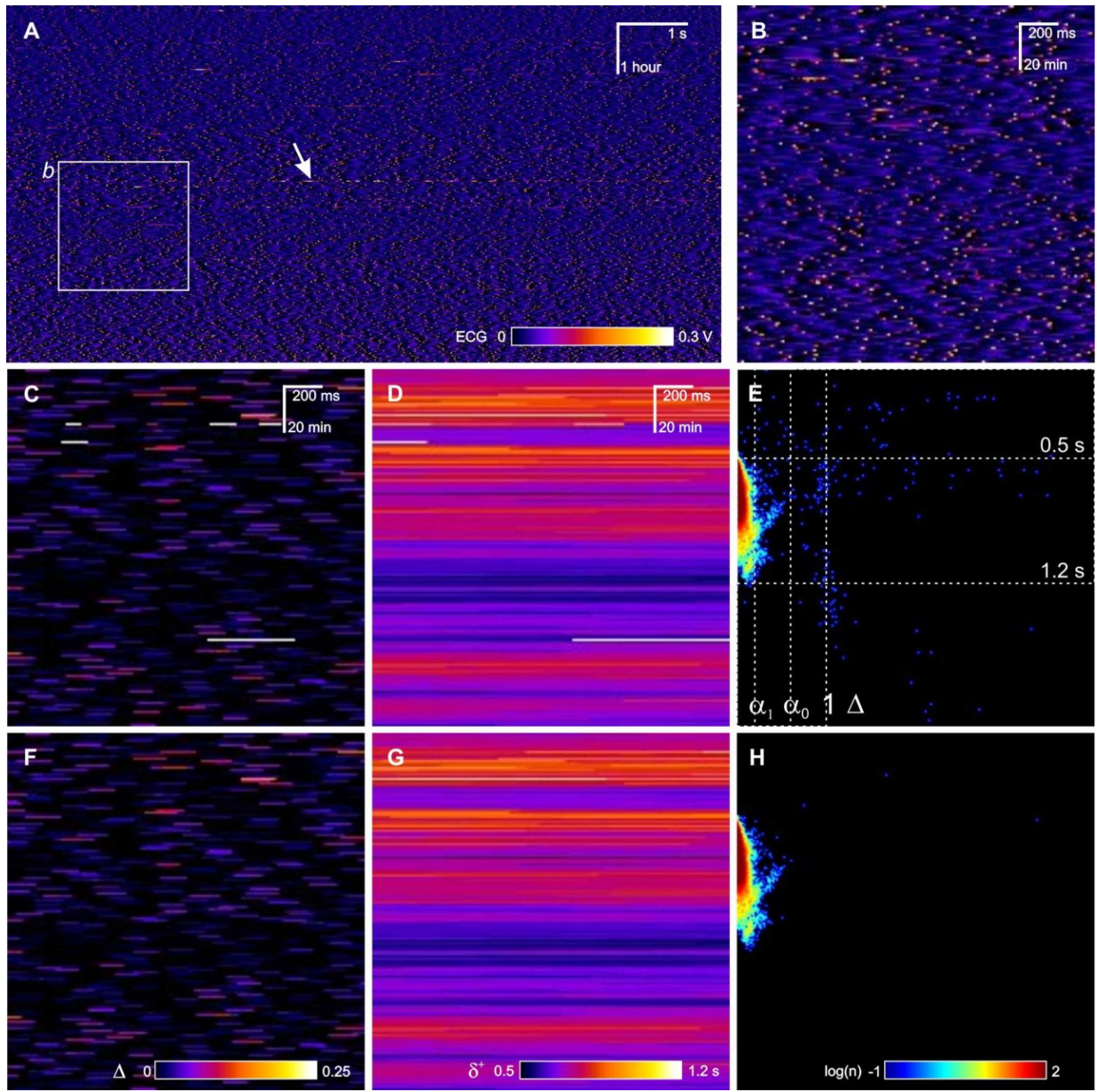

Figure 6 | R-peak detection in CardioImage

A, raster image of a seven hour ECG. Noisy segments are visible (*arrows*). B, zoomed in segment of the image. R-peaks are apparent as white specks. The ECG was run through the NN trained with false-positives (the ECG was not used to for any of the training data). C, Δ for the segment in B. To enhance visualization Δ is filled in only over the first fifth of the interval following the detection. D, δ⁺for the same segment. E, δ⁺/Δ co-distribution. The thresholds for alarm sequence detection, $α_0$ and $α_1$ (Methods) are indicated by white dashed lines, as well as Δ = 1, along which can be seen a cluster of detections. F- H, the same segments after action of alarm algorithm.

The mean and S.D. of the ECG channel (used to normalise the ECG for the NN) are calculated from a user-selected, noise-free region of its image. The NN then detects R-peaks in the ECG channel. For our model (Table 1) this took ~35 s for an eight-hour recording sampled at 500 Hz (14.4 x $10^6$ samples), using an Intel Core i5 processor and 12 GB of memory. Once finished, the detection channel is filled in - detections are marked by upward ticks of amplitude 1 (Fig. 5). The $\delta^+/\Delta$ channel (the user can switch between the alarm measures) provides a quick summary of the likely errors made by the NN (Fig. 6 C, D). Also the $\delta^+/\Delta$ co-distribution can be plotted as an 2-D histogram image. This clearly shows the integer clusters in $\Delta$ (Fig. 6 E).

The alarm algorithm is activated by the user and completes in milliseconds. The $\Delta$ integer peaks have been removed (compare Fig. 6 E and H). Alarm sequences longer than a selected threshold can be filled in with not-a-number (NaN) values, if it is felt that interpolation over these would introduce too much artificial data. Also the algorithm can be undone - interpolated and NN detections are marked with different amplitudes (0.5 and 1) in the detection image, allowing their unique identification.

The detection stack can be examined in a more usual, scrolling graphical display (Fig. 5). Using this display, mouse clicks can be used to manually add or remove individual detections and to interpolate or NaN segments. Our experience has been that the NN and alarm algorithm make only one or two mis-detections that need to be edited for every hour of recording. These were usually either premature contractions where a true detection has been removed (see above) or poor interpolations. The inspection and editing of an hour of ECG was usually complete in two minutes.

Once the user is finished editing, the detections channel is added to the original data stack and the detections stack is closed. This original stack, now with the addition of the detections, can then be saved as tiff file using the standard ImageJ interface.

3) *Tachograms*

From the detections channel the Intervals plugin outputs either a list of intervals and their times or a tachogram raster image (Fig. 7 A). For the image the tachogram is sampled at a fixed frequency using either linear, cubic or spline interpolation. Respiratory sinus arrhythmia and premature contractions, stand out clearly in these images (Fig. 7 D, E).

4) *Cycle Ensembles*

Ensemble (averaged) cycles can be calculated automatically for specified length windows (e.g. a minute). Ensembles from consecutive, overlapping windows with at a specified interval (e.g. 10 s) are then stacked into a cycle time versus time image (Fig. 7 B). Changes in PQRST morphology and the cycle morphology in any other channel (such as impedance), can be visually tracked over time. The averaging can be done directly on each cycle, or each cycle can be stretched to a unit standard before averaging.

5) *Spectral Analysis*

Frequency spectra analysis is central in traditional HRV (see Discussion). A plugin calculates Fourier spectra automatically for specified length windows (e.g. a minute). In the same fashion as the cycle ensembles, these spectra are calculated for consecutive, overlapping windows to create a frequency-time image. Such an image is commonly called the spectrogram (see Discussion). The spectral amplitude can be shown as absolute (ms), density (ms$^2$/Hz) or logarithmic (dB). Bands of interest such as the respiratory sinus arrythmia. can be easily followed over time (Fig. 7 C).

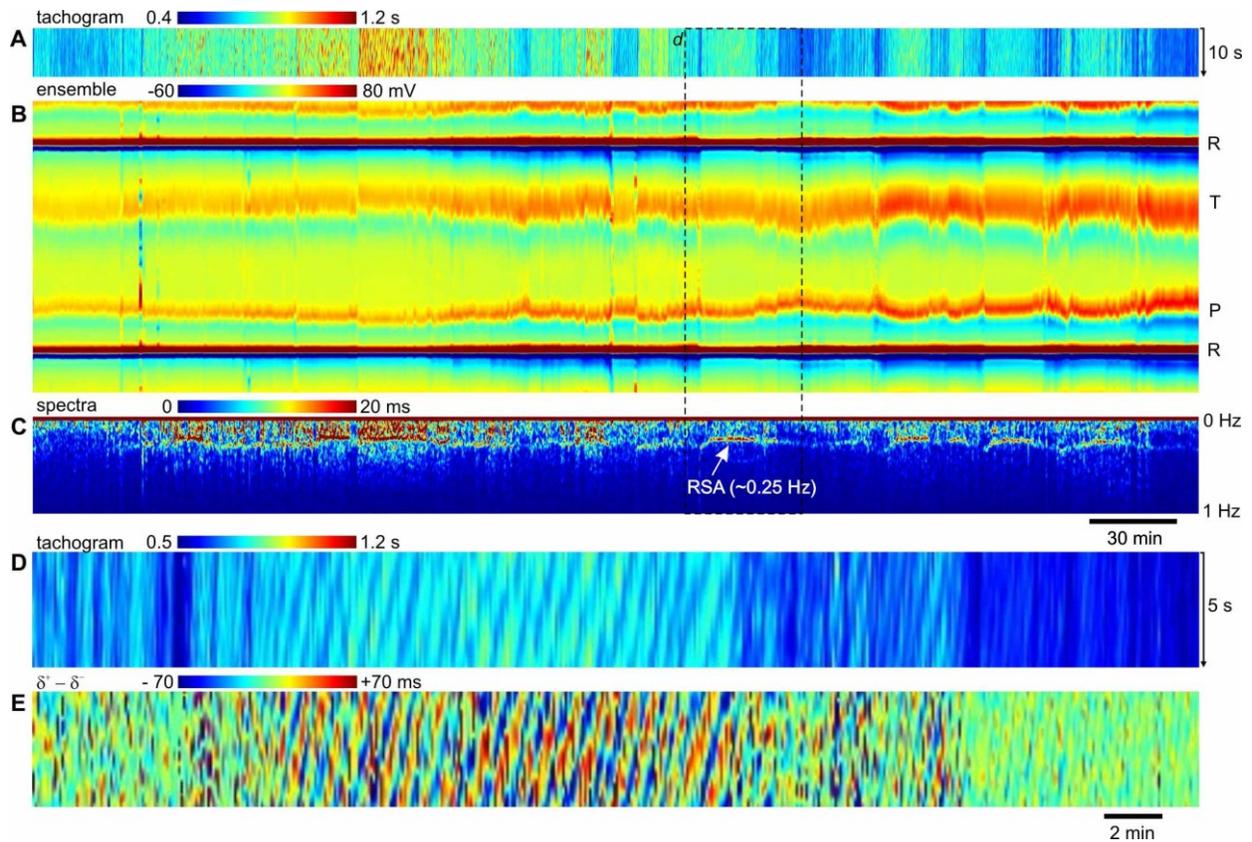

Figure 7 | Tachogram analysis in CardioImage

Corresponding tachogram (A), ensemble cycles (B) and Fourier frequency spectra (C) for a 400 minute recording. Time is from left to right. The tachogram was calculated by bicubic interpolation of RR intervals at 10 Hz sampling frequency and is arranged as a raster image with time running from top to bottom (10 s) and then left to right. The ensembles were calculated for minute length windows, overlapped at intervals of 10 s. Each RR cycle is resampled before averaging so that the RR interval covers a standard length - therefore the y-axis (cycle time running from top to bottom) has no real units. The P, R and T peaks are clearly visible (as indicated). The spectra were also calculated over minute length windows at intervals of 10 s. A band at ~0.25 Hz, corresponding to the respiratory sinus arrhythmia (RSA), appears intermittently. D, an expanded segment of the tachogram (indicated by dashed box) during one of these RSA bands, shows a stripe pattern as the RR interval regularly oscillates with a period ~ 4 s. This even clearer when the change in consecutive RR interval ($\delta^+$ - $\delta^-$) is plotted (E).

DISCUSSION

Here we described a pipeline for fast calculation of tachograms from noisy ECGs of several hours' length. The pipeline consists of three stages:

1) A neural network (NN) trained to detect R peaks and distinguish these from noise.
2) A measure to robustly detect false positives (FPs) and negatives (FNs) produced by the NN.

3) A simple "alarm" algorithm for automatically removing FPs and interpolating FNs.

In addition, we introduce the approach of encoding ECGs, tachograms and other cardiac time series in the form of raster images, which greatly speeds and eases their visual inspection and analysis.

*R detection: the importance of timing*

The performance of our NN, its ability to detect QRS embedded in noise, might be improved just by the increasing the amount of training data. We expect this would be minor. Another way to increase performance would be to give it another channel to work with. We always record impedance (a measure of blood volume in the aorta) alongside the ECG and as they vary on the same cycle the impedance might provide enough extra information to improve the detection. However the greatest increase in performance might result from introducing another dimension - timing.

When we manually edited noisy segments, we noticed that our judgment, our brain's detector, was based not just on waveform shape, but on timing. When a QRS wasn't apparent at first glance, we might look more closely at where (or more accurately when) we expected to see one in relation to those QRS that were immediately apparent. Such timing cues are absent in our NN and in the alarm algorithm they are present in a purely destructive way - we remove whole sequences of detections where the timing seems suspect and then fill in with artificial interpolates. In traditional signal-processing algorithms like the PT there is also usually a decision process based on timing - for instance a detection gets supressed if it is too soon after another (i.e. the interval is too short to be physiological).

One possible approach to introduce timing using NNs would be as follows:

1) Use the present NN to detect QRS base on morphology alone, but do not train it with a false positive bank. In fact maximise the sensitivity (detection of TPs) at the expense of selectivity (suppression of FPs), the opposite of our scheme.

2) Use the alarm algorithm to flag up segments likely to have FPs, but instead of removing all the detections over the alarm sequence, somehow do the following. Compare the activation signals of the logit neurons (R-peak and not-an-R-peak) to some model of the heart rate, perhaps based on data from clean sections of the recording. By this comparison estimate the most likely sequence of detections. We say, somehow, because as of yet we have no concrete idea how to do this. One possibility might be to use a recurrent NN (RNN) to build a model of the heart rate. In our NN the input layer takes a short segment (window) of the times series and this window is incremented (slid) across the time series to produce corresponding logit signals. In an RNN the same occurs, but each neuron takes as input, in addition to a sample in the current window, its activation by the previous sample in the previous window. In this way the RNN has a form of memory and can be used to predict what will come next in a time series - it forms a model of the times series. The RNN could be used in concert with the logit signals of the first NN (1) to estimate when a QRS likely did occur, but we do not know exactly how as yet. This approach, where the output of two independent algorithms are used to make an overall judgement, that is possibly more accurate then either alone, is called ensemble learning in the machine learning field.

Abrishami et al. (2018) applied a long short-term memory NN, a type of RNN, to peak detection in ECGs [12]. They used it for detection not just of QRS, but also P and T waves - they had four prediction classes, as opposed to our two (R-peak and not-an-R-peak). They obtained a QRS sensitivity of 95 % and selectivity of 94 % on apparently clean ECGs.

*Detecting Noise*

The alternative to detecting QRS in noise, is to exclude noisy segments of ECG from analysis entirely. To this end, two recent studies have looked to identify noisy segments with machine learning.

In machine learning the set of features that define some object is called its feature vector. Each feature must be a numerical value or something that can be coded as a numerical value. For instance, the feature vector for a piece of fruit might be its weight, circumference and color, but the latter would have to coded numerically (0 = read, 1 = green, etc). Each of the $n$ features of an object can be thought of as an axis in an $n$-dimensional space. A feature vector (object) is a point in this space, its coordinates the value of each feature. Two common tasks in machine learning are:

1) *Classification*. Divide the feature space into non-overlapping volumes, each of which has a class. A feature (object) can then be assigned to a class by its coordinate (feature values). Our NN is a classifier - it classified each window of samples (feature vector) as either "R-peak" or "not an R peak". However the classes may just as well have been "noisy" and "not noisy". This is in fact what John et al. (2018) did, using a CNN and non-overlapping windows of 1 to 30 s length [28].

2) *Clustering*. Group objects into clusters in feature space - groups of points that lie near to each other. In agglomerative clustering the algorithm starts by defining each feature vector as a cluster, then clusters those clusters, and so on iteratively, until there is just one cluster left containing all the vectors. Rodrigues et al. (2017) used agglomerative clustering for noise segmentation. The feature vector consisted of a window of samples (as for us and [28]), plus some summary statistics for that window (standard deviation, range, etc). They found that of the top three or four clusters in the hierarchy, one of these would accord with noise [29].

The disadvantage of clustering, as opposed to a classification algorithm, is that noise is defined by the clustering of the input windows themselves, rather than being an *a priori* class to which a window is assigned. The clustering has to be run anew on each ECG you want to segment (which might involve more computation than running it through an already trained neural net). Then the noise cluster has to be identified and the definition of that cluster may shift with the data itself. Our alarm algorithm segments noise *a priori* by detecting unphysiological values of Δ resulting from misdetections (FPs and FNs) by the NN.

*Image based Analysis and HRV*

The advantage of displaying time series as images with a time axis, is that the information in them is compressed into a short space and can be visually digested at a glance. As far as we are aware we are the only study to have applied this to the tachogram itself and ensemble cycles. One study plotted consecutive cycles overlapping, together with their average [30].

The practice of plotting the frequency spectra of a time series as an image, with time along one axis and frequency along the other, is ubiquitous across science and engineering with applications from the analysis of bird song to the detection of gravitational waves [31, 32]. The images have variously been called spectrograms, spectral waterfalls or sonograms. The first spectrograms were created by analogue machines during world war II for naval and cryptographic uses [32-34]. They became even more popular in the digital era thanks to the invention of the fast Fourier transform in 1965. Most recently they have seen a resurgence with the invention of wavelet techniques. Spectral analysis of tachograms is central to HRV, so it is surprising that only a scattering of publications over the years have used spectrograms [e.g. 35, 36]. Most publications only give the integral of a particular band for a time period of interest. In some this is supplemented by the spectra for a single time window [37]. This is a great pity given the richness of information provided by the spectrogram (Fig. 7 C). It probably results from the reliance of most authors on commercial ECG software for analysis - there are maybe just a couple that implement spectrograms (Autonom Health, Medilog by Schiller).

*Summary*

QRS detection is a time consuming task with current commercial software. The PT-type algorithms they use mean that mis-detections are inevitable for ECGs of more than a few minutes and manually editing these can take a very long time in our experience. For HRV to realise its full potential it is paramount that QRS detection is improved by bringing it into the era of artificial intelligence. Also to understand heart rate variability as a physiological phenomenon, we must go beyond defining "HRV" as just a collection of singular measurements (HF/LF ratios, RMSD, etc) and look more closely at temporal patterns. Images such as the spectrogram and ensemble series, can only help in this.